\documentstyle[11pt]{article}
\catcode`\@=11
\begin{document} \openup6pt

\title{  Modified Chaplygin Gas as Scalar Field \\
and Holographic Dark Energy Model
}

\author{B. C. Paul\thanks{Electronic mail : bcpaul@iucaa.ernet.in}  \\
    Physics Department, North Bengal University, \\
Siliguri, Dist. : Darjeeling, Pin : 734 013, West Bengal, India \\
P. Thakur \\
Physics Department, Alipurduar College, Dist. : Jalpaiguri, India
\\
A. Saha\thanks{ Electronic mail : arindamjal@gmail.com}\\
Physics Department, Darjeeling Government College, Dist. :
Darjeeling, India}

\date{}

\maketitle

\vspace{0.5in}

\begin{abstract}

We study the correspondence between field theoretic and holographic
dark energy density of the universe with  the modified  Chaplygin
gas (MCG) respectively both in a flat and non-flat FRW universe. We
present an equivalent representation of the MCG with a homogeneous
minimally coupled  scalar field by constructing the corresponding
potential. A new scalar field potential is obtained here which is
physically realistic and important for cosmological model building.
In addition we also present  holographic dark energy  model
described by the MCG. The dynamics of the corresponding holographic
dark energy field is determined by reconstructing the potential in a
non-flat universe. The stability of the holographic dark energy in
this case in a non-flat universe is also discussed.

\end{abstract}

\vspace{0.2cm}

PACS number(s) : 04.20.Jb, 98.80.Cq, 98.80.-k

\vspace{4.5cm}

\pagebreak

\section{ Introduction:}

In the recent years a number of observations have been carried 
out that lead to a precise knowledge
of the cosmological evolution. Cosmological observations like abundances of galaxy
cluster large scale redshift surveys [1], angular power spectrum of CMBR [2],
and baryon oscilllations [3] suggest that the universe is nearly flat and almost 73 \% of the matter
in the form of dark energy. Further the magnitude 
redshift surveys of type Ia Supernovae indicates the  universe has
recently entered a phase of accelerating
phase of expansion [4]. On the otherhand it is generally accepted that our universe
might have also emerged from an accelerating phase in the past. Thus one of the essential
ingradients in  modern cosmology is inflation.
However, inflation cannot be accommodated in a perfect fluid assumption in
the framework of Einstein general theory of Relativity (GTR). This led to explore
gravitational theories with a  modification of the matter sector or
the gravitational sector. 
It is known that  the early inflation may be realized in a semiclassical theory
of gravity where matter is described by quantum fields [5].
Starobinsky also obtained inflationary solution 
considering a curvature squared term in the Einstein-Hilbert action
[6] long before the advent of inflation was known. However, the
efficacy of inflation is known only after the seminal work of Guth
who first employed the phase transition mechanism to accommodate
inflation.  A  large volume  of
literature appeared with a modification of gravitational sector in
which curvature squared terms [7] were added to the Einstein-Hilbert
action to realize early inflationary universe scenario. In the case of 
inflaton field cosmology early inflation  may be realized  with an 
equation of state $p = \omega \; \rho$, where $\omega = -1$. 
However, an accelerating late universe with barotropic fluid emerges 
when $\omega < -1$.  The usual matter
fields in the standard model of particle physics unable to
accommodate the late accelerating phase of the universe. Therefore,
it is a challenging task  to formulate a consistent theoretical
framework which might accommodate the  observational facts. Recent
astronomical data when interpreted in the context of Bigbang model
have provided some interesting information about the composition of
the universe. It is beleived that the universe is dominated by a huge amount of dark energy.
 To accommodate such a huge energy  various  mechanism have been proposed
 during the last
 few years in order to classify the physical nature of the cosmic fluid [8].
Among the different theories proposed, the single component fluid
known as Chaplygin gas [9]
 with  an equation of state $p = - \frac{A}{\rho}$, where $\rho$ and $p$
 are the energy density and
 pressure respectively and $A$ is a constant, has attracted large interest
 in cosmology [7].
 The above equation of state,
 however, has been conceived in studies of adiabatic  fields. It was
 used to describe lifting forces on a plane wing in aerodynamic process.
 In cosmology,  although it admits an  accelerating universe [10],  fails
 to address structure formation and cosmological perturbation power
 spectrum [11].
 Subsequently, a generalized  form of
 the equation of state (in short, EOS)
 $p = - \frac{A}{\rho^{\alpha}}$ with $0 \leq \alpha \leq 1$
 was considered to construct a viable
 cosmological model [12, 13],    which is  known as generalized
 Chaplygin gas (in short, GCG) in cosmology. It has two free parameters
 $A$ and $\alpha$.
 The fluid behaves initially like  dust for small size of the universe.
 but at a later epoch the fluid  may be described by  an equation of state
  $p = \omega \rho$.
   It has string connection, the above equation of state can be obtained from the
   Nambu-Goto  action for a D-brane moving in a (D+2)-dimensional space-time
   in the
   light cone parametrization [14]. Recently a new  form of equation of state $p =
 f(\rho)$ has been considered to study  the dark energy content of the
 universe [15]. Consequently a three parameter modified form
   of the equation of state
   for Chaplygin gas is more important. Therefore, we consider
   an equation of state of the form
 \begin{equation}
p = B \rho - \frac{A}{\rho^{\alpha}} \; \;  \; with \; \; 0 \leq
\alpha \leq 1,
\end{equation}
where $B$ is an equation of state parameter and $ A $ is a constant,
which is termed as modified Chaplygin  gas (in short, MCG) [16]. An
interesting feature of the MCG equation of state is that in the
early universe when the size of the universe $a(t)$ was small, it
behaved like a barotropic fluid ( if one considers $B = \frac{1}{3} $
it corresponds to radiation and $B=0$ it corresponds to matter ) but
at a later epoch it behaves as a cosmological constant which can be
fitted to a $\Lambda$CDM model. Recently, the thermal equation of
state of the MCG is studied [17] and it is found that the MCG may
cool down through  thermodynamic processes without facing any
critical point or phase transition. They noted the following
constraints  for a realistic solution on the values of the parameters : (i) for $B \sim 0$,
$0 < \alpha < 1$, (ii) for $B = \frac{1}{3}$, $0 < \alpha <
\frac{1}{2}$, (iii) for $B = 1$, $\alpha \sim 0$. Barrow [18] has
outlined a method  to fit
 the Chaplygin gas in the FRW universe.
In a flat Friedmann model it is shown [10]  that the generalized
Chaplygin gas may be equivalently described in terms of
 a homogeneous minimally coupled scalar field $\phi$.  Gorini {\it  et al.} 
 [19] using the above scheme obtained
 the corresponding
homogeneous scalar field $\phi (t)$   and the corresponding
potential $V(\phi)$ which can be used to obtain a viable
cosmological model with the generalized Chaplygin gas. Thus it is
important to look for the field and the relevant potential for the
modified Chaplygin gas, which will be discussed in the section 3.

Recently, another interesting topic, namely holographic principle
[20, 21] is incorporated in cosmology [22-25] to track the dark
energy content of the universe following the work of Cohen {\it et
al.} [26]. Holographic principle is a
 speculative conjecture about quantum gravity theories proposed by G't Hooft [27].
 The idea has been subsequently
 promoted by Susskind and his coworkers [20] claiming that all the information
 contained in a spatial volume may
  be represented by a theory that lives on the boundary of that space.
For a given finite region of space it may  contain matter and energy
within it. If this energy  suppresses a critical   density then the
region collapses to a black hole. A black hole is known
theoretically to have an entropy which is proportional to its
surface area of its event horizon. A black hole event horizon
encloses a volume, thus a  more massive  black hole have larger
event horizon and encloses larger volume. The most massive black
hole that can fit in a given region is the one whose event horizon
corresponds exactly to the boundary of the given region under
consideration. The maximal limit of entropy for an ordinary region
of space is directly proportional to the surface area of the region
and not to its volume. Thus, according to holographic principle,
under  suitable conditions all the information about a physical
system inside a spatial region is encoded in the boundary. The basic
idea of a holographic dark energy in cosmology is that
 the saturation of the entropy bound may be related to an  unknown ultra-violet (UV)  scale
$\Lambda$ to some known comological scale
  in order to enable it to find a viable formula for the dark energy which may be quantum gravity in
  origin and it is  characterized by $\Lambda$. The choice of UV-Infra Red (IR) connection from the covariant
  entropy bound leads to a universe dominated by blackhole states.
According to   Cohen {\it et al.} [26] for   any state in the
Hilbert space with energy $E$, the  corresponding  Schwarzschild
radius $R_s \sim E$, may be less than the IR cut off value $L$
(where $L$ is a cosmological scale). It is possible to derive a
relation between the UV  cutoff $\rho_{\Lambda}^{1/4}$ and the IR
cutoff which eventually
   leads to a constraint   $ \left( \frac{8 \pi G}{c^2} \right)
   L^3  \left( \frac{\rho_{\Lambda}}{3}
   \right) \leq L$ [26] where $\rho_{\Lambda}$ is the energy density corresponding to dark energy
   characterized by $  \Lambda$, $G$ is Newton's gravitational constant  and $c$ is a parameter in the theory.
   The holographic dark energy density  is
\begin{equation}
\rho_{\Lambda} = 3 c^2 M_P^2 L^{-2},
\end{equation}
where $M_P^{-2} = 8 \pi G$. It is known that the present
acceleration may be described if $\omega_{\Lambda}
=\frac{p_{\Lambda}}{\rho_{\Lambda}} < - \frac{1}{3}$. If one
considers $L \sim \frac{1}{H}$ it gives  $\omega_{\Lambda} =0$. A
holographic cosmological constant model based on Hubble scale as IR
cut off does not permit accelerating universe. It is also  examined
[22]  that the holographic dark energy model based on  the particle
horizon as the IR cutoff even does not work to get an accelerating
universe. An alternative model of dark energy  using particle
horizon in closed model is also proposed [26]. Recently, Li [23] has
obtained an accelerating universe considering  event horizon  as the
cosmological scale. The model is consistent with the cosmological
observations.  Thus to have a model consistent with observed
universe one should adopt the covaiant entropy bound and choose $L$
to be the event horizon.

The motivation of the paper is two folds (i) to explore an
equivalent scalar field representation corresponding to
the MCG in cosmology and (ii) to explore an equivalent holographic
dark energy field, considering the event horizon as the cosmological
scale in a non-flat universe. We obtain holographic description of the MCG 
dark energy in FRW universe and reconstruct the potential and the 
dynamics of the scalar field which describes the MCG cosmology. We consider
both flat and  non-flat universe here as it is not yet decided.
The paper is organized as follows : in sec. 2, the
relevant field equation with modified Chaplygin gas in FRW universe
is presented; in sec. 3 we present an equivalent field theoretic representation of MCG
by a scalar
field,  constructing the corresponding potential, in sec. 4, we suggest a correspondence
between holographic dark energy fields with MCG.  In
sec. 5, squared speed of sound for holographic dark energy is
evaluated for a closed universe i.e., $k = 1$ to study the stability
of the field. Finally in sec. 6, a brief discussion.

\section{ Modified Chaplygin Gas in FRW universe :}

The Einstein's field equation is
given by
\begin{equation}
R_{\mu \nu} - \frac{1}{2} g_{\mu \nu}  =  \kappa^2 \; T_{\mu \nu}
\end{equation}
where $\kappa^2 = 8 \pi G$,  and $T_{\mu \nu}$ is the energy
momentum tensor.

We consider a homogeneous and isotropic universe given  by
\begin{equation}
ds^{2} = - dt^{2} + a^{2}(t) \left[ \frac{dr^{2}}{1- k r^2} + r^2 ( d\theta^{2} + sin^{2} \theta \;
d  \phi^{2} ) \right]
\end{equation}
where $a(t)$ is the scale factor of the universe, the matter is
described by the energy momentum tensor $T^{\mu}_{\nu} = ( \rho, p,
p, p)$ where  $\rho$ and $p$ are energy density and pressure
respectively.

Using the metric (4) and  the energy momentum tensor, the  Einstein's field equation (3)  can be written as
\begin{equation}
H^{2}+\frac{k}{a^2} =  \frac{1}{ 3 M_{P}^2 } \rho
\end{equation}
where we use $8 \pi G = M_{P}^{-2} = \kappa^2$. The conservation
equation for matter is given by
\begin{equation}
\frac{d\rho}{dt} + 3 H (\rho + p) = 0 .
\end{equation}
For  the  modified Chaplygin gas, using the EOS given by eq. (1), the eq. (6) can be
integrated to obtain the energy density which is given by
\begin{equation}
\rho = \left(  \frac{A}{1+B} + \frac{C}{a^n} \right)^{\frac{1}{1+
\alpha}}
\end{equation}
where  $C$ is an arbitrary  constant and we denote $  3(1 + B) ( 1 +
\alpha) = n$. We  define the following
\begin{equation}
\Omega_{\Lambda} = \frac{\rho_{\Lambda}}{\rho_{cr}}, \; \Omega_{m} = \frac{\rho_{m}}{\rho_{cr}}, \;
\Omega_{k} = \frac{k}{a^2 H^2}
\end{equation}
where $\rho_{cr} = 3 M_P^2 H^2$, $\Omega_{\Lambda}$, $ \Omega_m $
and $\Omega_k$ represent  density parameter corresponding to
$\Lambda$, matter and curvature respectively in the paper.

\section{Modified Chaplygin Gas (MCG) as a scalar field :}

In this section, we obtain the field theoretic represntation of 
the modified Chaplygin gas assuming a homogeneous
scalar field $\phi (t)$. We use Barrow's scheme [18] here and the corresponding
 energy density and pressure of the homogeneos field are identified as:
\begin{equation}
\rho_{\phi} = \frac{1}{2} \dot{\phi}^2 + V (\phi) = \left(
\frac{B}{A+1} +  \frac{C}{a^n} \right)^{\frac{1}{\alpha +1}},
\end{equation}
\begin{equation}
p_{\phi} = \frac{1}{2} \dot{\phi}^2 - V (\phi) = \frac{ - \frac{
B}{A + 1} +  A \frac{C}{a^n}}{
 \left( \frac{ B}{A+1} +  \frac{C}{a^n} \right)^{\frac{\alpha}{\alpha
 +1}}}.
\end{equation}
which are equated with that of MCG given by eqs. (1) and (7)
respectively. The scalar field potential
and the corresponding kinetic energy  of the field are obtained 
 from eqs. (9) and (10), which are
\begin{equation}
V(\phi) = \frac{  \frac{A}{B + 1} + \frac{1 - B}{2}  \frac{C}{
a^n}}{ \left( \frac{ A}{B+1} +  \frac{C}{a^n}
\right)^{\frac{\alpha}{\alpha +1}}},
\end{equation}
\begin{equation}
\dot{\phi}^2 = \frac{ (B + 1) \frac{ C}{ a^n}}{  \left( \frac{
A}{B+1} +  \frac{C}{a^n} \right)^{\frac{\alpha}{\alpha +1}}}.
\end{equation}
In the next section we consider the above equations to determine the 
potential in the two cases (i) flat universe and (ii)
non-flat universe using eq. (5).

 Case I :  For a flat universe ($k = 0$),  using
eq. (5)  and (9) we get
\begin{equation}
H^{2} =  \frac{1}{ 3 M_{P}^2 } \left( \frac{B}{A+1} + \frac{C}{a^n} \right)^{\frac{1}{\alpha +1}}.
\end{equation}
Using eqs. (12) and (13) we get
\begin{equation}
\phi - \phi_o = \pm \frac{2 M_P}{\sqrt{n (1 + \alpha)}} \; sinh^{-1}
\left[ \sqrt{ \frac{C (B + 1)}{A}} a^{- \frac{n}{2}} \right]
\end{equation}
and the  corresponding potential is
\begin{equation}
V(\phi) = \frac{ \frac{A}{1 + B} + \frac {A (1 - B) }{2 (1 + B)} \;
sinh^2 \left( \frac{ \sqrt{\frac{n(1+\alpha)}{2M_P}}} \; (\phi - \phi_o) \right) }{
\left( \frac{A}{1 + B} \right)^{\frac{\alpha}{\alpha + 1}}  \;
cosh^{\frac{2 \alpha}{1 + \alpha}}  \left( \frac{ \sqrt{\frac{n(1+\alpha}{2M_P}}}\;
(\phi - \phi_o) \right)}.
\end{equation}

The potential asymptotically approaches a constant as $\phi \rightarrow \phi_o$,
however, it increases with increasing value of the field if $\phi \neq \phi_o$. The
potential obtained by Gorini {\it et al.} [19] can be recovered here
if one puts $\alpha = 1$ and $B = 0$.

Case II : For a non flat ($ k \neq 0$) universe,  the
evolution of the scalar field is obtained  using eqs (5), (9) and (12),
which is
\begin{equation} \phi - \phi_o = \pm \int \sqrt{ \frac{12
M_P^{2} (B + 1)}{n^2}} \; \frac{dz}{\sqrt{\left( \mu^2+z^2 \right)
-\frac{3 M_P^{2} k}{C^{2/n}} \; z^{4/n}
(\mu^2+z^2)^{\frac{\alpha}{\alpha +1}}}}
\end{equation}
where $k = +1$ for closed universe ( $k = -1$ for open universe )
and we denote $\mu^2= \frac{A}{B+1}$, $z= \sqrt{\frac{c}{a^n}}$. The
above integration is not simple so as  to express the potential in terms of
the field. However,  for some special choice of the parameters
the potential may be obtained which can be expressed in terms
of the field $\phi$. We choose the following :

$\bullet$ $B = - \frac{1}{3}, \; \; A = 0$, the scalar field evolves
as
\begin{equation}
\phi_{\pm}  = \phi_o \pm \sqrt{ \frac{ 2 M_P^2}{n^2(1 - \frac{3
M_P^2 k}{c^{2/n} })}} \; ln \; \left( \frac{C}{a^n} \right),
\end{equation}
the corresponding scalar field potential is given by
\begin{equation}
V(\phi) = \frac{2}{3}  \; Exp \left[
\frac{1}{ \sqrt{ \frac{2 M_P^2 (\alpha +1)^2}{n^2(1- \frac{3M_P^2
k}{c^{2/n}} )} } \; (\phi - \phi_o)} \right].
\end{equation}
It is an  exponential potential which increases  (
decreases ) depending on the evolutionary behaviour of the scalar field  $\phi_{+}$
( $\phi_{-}$  ) as given in (17). We  note that the potential is
positive definite if  $C
> \left( 3 M_P^2 k \right)^{n/2}$. In the case of closed universe
the above inequality gives a lower bound on the values of $C$ which
is a positive number. But for an open universe $C$ picks up both positive and
negative values   for an even integral values of $n$.

 $\bullet$ $B = \frac{1}{3}, \; \; A \neq 0$, in
this case the scalar field evolves as
\begin{equation}
\phi - \phi_o = \pm \sqrt{ \frac{ 16 M_P^2}{n^2(1 - \frac{3 M_P^2
k}{c^{2/n}})}} \; sinh^{-1}\; \left(\frac{4C}{3A} \frac{1}{a^{n/2}}
\right),
\end{equation}
and the corresponding potential is given by
\begin{equation}
V(\phi) = sech^2 \; \left( \frac{n^2(1- \frac{3M_P^2 k}{c^{2/n}}
)}{16 M_P^2} \right)  \; (\phi - \phi_o)  + \frac{1}{3} \; tanh^2
\;\left( \frac{n^2(1- \frac{3M_P^2 k}{c^{2/n}} )}{16 M_P^2} \right)
\; ( \phi - \phi_o).
\end{equation}
The  fig. 1 is a plot of the potential in terms of the field. It may be noted
that in the case of a flat universe ($k =0$), one obtains a potential
different from that obtained in {\it Ref. } [19] as $B \neq 0$.
 It is a new and interesting potential, which has a shape
similar to that one obtains in the case of tachyonic field [29]. 
The potential attains to a constant value at a large time  leading to
 late acceleration of the universe.   We also note that  an oscillatory scalar field results for $C <
\left( 3 M_P^2 k \right)^{n/2}$ in  a sinusoidal
potential in a closed universe. However, in the case of an open universe a realistic solution is permitted for an 
even integer
values of $n$ only.
\input{epsf}
\begin{figure}
\epsffile{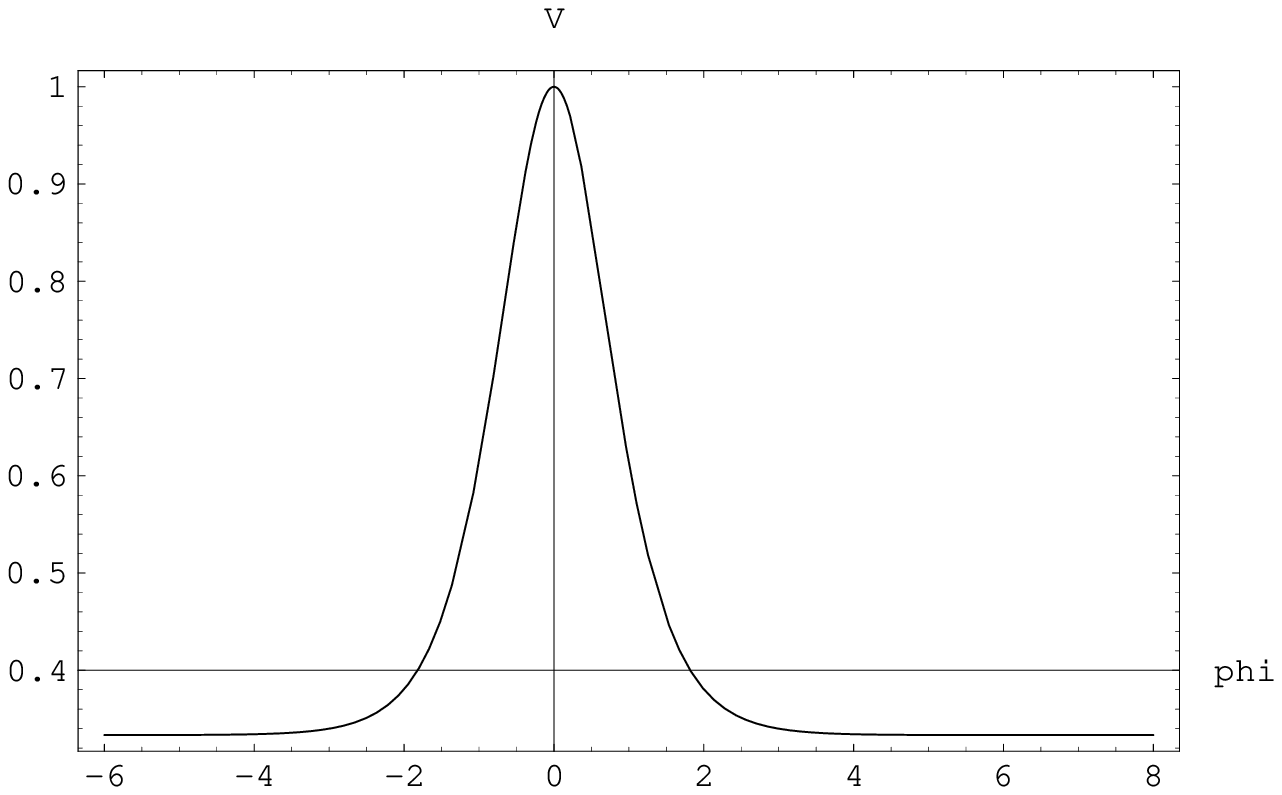} \caption{shows the plot of $V$ versus $\phi$
with the parameter $ \frac{n^2(1- \frac{3M_P^2 k}{c^{2/n}} )}{16
M_P^2} =1$.}
\end{figure}

\section{MCG as Holographic Dark Energy  :}

In a FRW universe we now consider a non-flat universe with $k \neq 0$
 and use the holographic dark energy density as given in (2) which is
\begin{equation}
\rho_{\Lambda} = 3 c^2 M_P^2 L^{-2},
\end{equation}
where $L$ is the cosmological length scale for tracking the field
corresponding to holographic dark energy in the universe and $c$ is
a parameter. The parameter $L$ is defined as
\begin{equation}
L = a r (t).
\end{equation}
where $a(t)$ is the scale factor of the universe and $r(t)$ is
relevant to the future event horizon of the universe. Using
Robertson-Walker metric one gets [24]
\[
L = \frac{a (t)}{\sqrt{|k|}}  \;  sin \; \left[ \sqrt{|k|}
R_{h}(t)/a(t) \right] \;\; for \; \; \; k = +1 ,
\]
\[
\; \; \; =   R_h \;\; for \; \; k =0,
\]
\begin{equation}
 \; \; \; =  \frac{a (t)}{\sqrt{|k|}} \; \; sinh \; \left[
\sqrt{|k|} R_{h}(t)/a(t) \right]\; \; for \; \; k = - 1 .
\end{equation}
where $R_{h}$ represents the event horizon which is given by
\begin{equation}
R_h = a(t) \; \int_t^\infty \frac{dt'}{a(t')} = a(t) \; \int^{r_1}_o \frac{dr}{\sqrt{1 - k r^2}}.
\end{equation}
 Here $R_h$ is
 measured in $r$ direction and $L$ represents the radius of
 the event horizon measured on the sphere of the horizon. Using
  the definition of $\Omega_{\Lambda} = \frac{\rho_{\Lambda}}{\rho_{cr}} $
  and $\rho_{cr} = 3 M_{P}^2 H^2$, one can derive  [25]
\begin{equation}
H L = \frac{c}{\sqrt{\Omega_{\Lambda}}}.
\end{equation}
Using eqs. (23)- (24), one determines the temporal rate of change of
$L$ which is
\[
\dot{ L} = \frac{c}{\sqrt{\Omega_{\Lambda}}} - \frac{1}{\sqrt{|k|}}
\; cos \; \left( \frac{\sqrt{|k|} \; R_h}{a(t)} \right) \; \; for \;
\; k = +1,
\]
\[
\; \; = \frac{c}{\sqrt{\Omega_{\Lambda}}} - 1 \; \; for \; \; k = 0,
\]
\begin{equation}
\; \; \; = \frac{c}{\sqrt{\Omega_{\Lambda}}} - \frac{1}{\sqrt{|k|}}
\; cosh \; \left( \frac{\sqrt{|k|} \; R_h}{a(t)} \right) \; \; for
\; \; k = - 1.
\end{equation}
Using eqs. (21) -(26) , it is possible to construct the required
equation for the holographic energy density $\rho_{\Lambda}$, which
is given by
\begin{equation}
\frac{d\rho_{\Lambda}}{dt} =  - 2 H \left[ 1 -
\frac{\sqrt{\Omega_{\Lambda}}}{c}  \; f(X) \right]
 \; \rho_{\Lambda},
\end{equation}
where we use the notation,
\begin{equation}
 f(X) = \frac{1}{\sqrt{|k|}} \; cosn \left( \sqrt{|k|} \; x \right) =
 cos (X) \; \left[ 1, cosh (X) \right]  \; fo r\;   k =1 \; [0, -1],
\end{equation}
where $X = \frac{R_h}{a(t)}$.
 The energy conservation equation is
\begin{equation}
\frac{d\rho_{\Lambda}}{dt} + 3   H (1 + \omega_{\Lambda})  \rho_{\Lambda} = 0
\end{equation}
which is used to determine the the equation of state parameter
\begin{equation}
\omega_{\Lambda}  =  - \left( \frac{1}{3} +  \frac{2
\sqrt{\Omega_{\Lambda}}}{3c} f(X) \right).
\end{equation}
Now  we consider the  holographic dark energy density which is assumed
to be equivalent to the modified  Chaplygin gas energy density. The
corresponding energy density is taken from (7), which is obtained
using the equation of state given by  (1). The equation of state
parameter corresponding to EOS (1) can be written as :
\begin{equation}
\omega  = \frac{p}{\rho} =  B  -   \frac{A}{\rho^{\alpha +1}}.
\end{equation}
Let us now establish the correspondence between the holographic dark
energy and modified Chaplygin gas energy density. In this case from
eqs. (7) and (21), we get
\begin{equation}
C = a^n \left[(3 c^2 M_P^2 L^{-2})^{1+\alpha} - \frac{A}{B+1}
\right].
\end{equation}
Now using eqs.  (30)-(32),  we determine the parameters as
\begin{equation}
A = ( 3 c^2 M_P^2 L^{-2}) ^{\alpha +1} \; \left[ B + \frac{1}{3} +
\frac{2  \sqrt{\Omega_{\Lambda}}}{3  c}  f(X) \right],
\end{equation}
\begin{equation}
C = ( 3 c^2 M_P^2 L^{-2})^{\alpha +1} \; a^n \left[ 1 - \frac{3 B
+1}{3(B + 1)} - \frac{2 \sqrt{\Omega_{\Lambda}}}{3 (B + 1) c} f(X)
\right].
\end{equation}
 The corresponding potential for the holographic dark energy  field  becomes
\begin{equation}
V( \phi)  = 2  c^2 M_P^2 L^{-2} \left[ 1 +  \frac{
\sqrt{\Omega_{\Lambda}}}{2 c} f(X) \right],
\end{equation}
and the corresponding kinetic energy of the field is given by
\begin{equation}
\dot{ \phi}^2  = 2  c^2 M_P^2 L^{-2} \left[ 1 -  \frac{
\sqrt{\Omega_{ \Lambda}}}{  c} f(X) \right].
\end{equation}
Considering $x$ $(= \ln a)$, we transform  the time derivative to the 
derivative with logarithm of the  scale factor,  which is the most 
useful function in this case. We get
\begin{equation}
\phi'= M_P \sqrt{ 2 \Omega_{\Lambda} \left( 1 - \frac{
\sqrt{\Omega_{\Lambda}}}{  c} f(X) \right)}
\end{equation}
where $()'$ prime represents derivative with respect to $x$. On integrating the above 
equation the evolution of the scalar field corresponding to holographic dark
energy is evaluated  which is given by
\begin{equation}
\phi (a) - \phi ( a_o) = \sqrt{2} M_P \int_{\ln a_o}^{\ln a} \sqrt{
\Omega_{\Lambda} \left( 1 -  \frac{ \sqrt{\Omega_{\Lambda}}}{  c}
f(X) \right)} \;  dx  .
\end{equation}
The above result is obtained for a non-flat universe ($k \neq 0$).
The flat case is not considered here as the holographic dark energy
in a flat FRW universe is unstable [30]. We study the stability of
the holographic dark energy model for a non-flat universe by
calculating the squared speed  in the next section.

\section{Squared speed for  Holographic Dark Energy :}

We consider a closed  universe model ($k = 1$) in this case. The
dark energy equation of state parameter given by eq. (30) reduces to
\begin{equation}
\omega_{\Lambda}  = - \frac{1}{3} \left( 1 + \frac{2}{c}
\sqrt{\Omega_{\Lambda}} \; cos \; y \right)
\end{equation}
where  $y = \frac{R_H}{a(t)}$. The minimum value it can take is
 $\omega_{min} = - \frac{1}{3}  \left( 1 + 2 \sqrt{\Omega_{\Lambda}}
  \right)$ and one obtains
 a lower bound $\omega_{min} = - 0.9154$ for
 $\Omega_{\Lambda}= 0.76$ with $c = 1$.
Taking variation of the state parameter with respect to $x = \ln \;
a$, we get [25]
\begin{equation}
\frac{\Omega_{\Lambda}'}{\Omega_{\Lambda}^2}= (1 - \Omega_{\Lambda})
\left( \frac{2}{c} \frac{1}{\Omega_{\Lambda}} cos \; y + \frac{1}{1
- a \gamma} \frac{1}{\Omega_{\Lambda}} \right)
\end{equation}
and the variation of equation of state parameter becomes
\begin{equation}
\omega_{\Lambda}' = - \frac{\sqrt{\Omega_{\Lambda}}}{3 c} \left[
\frac{1 - \Omega_{\Lambda}}{1 - \gamma a} + \frac{ 2
\sqrt{\Omega_{\Lambda}}}{c} \left(1 - \Omega_{\Lambda} cos^2
y\right) \right],
\end{equation}
where $\gamma = \frac{\Omega^{o}_k}{\Omega^{o}_m}$. We now introduce
the squared speed of holographic dark energy fluid  as
\begin{equation}
{\it v}_{\Lambda}^2 = \frac{dp_{\Lambda}}{d \rho_{\Lambda}} = \frac{\dot{p}_{\Lambda}}{\dot{\rho}_{\Lambda}} = \frac{p'_{\Lambda}}{\rho'_{\Lambda}},
\end{equation}
where  varaiation of eq. (31) w.r.t.  $x$ is  given by
\begin{equation}
p'_{\Lambda} = \omega'_{\Lambda} \rho_{\Lambda}+ \omega_{\Lambda} \rho'_{\Lambda}.
\end{equation}
Using the eqs. (42) and (43) we get
\[
{\it v}_{\Lambda}^2 = \omega'_{\Lambda} \frac{{\rho}_{\Lambda}}{{\rho'}_{\Lambda}} + \omega_{\Lambda}
\]
which now becomes
\begin{equation}
{\it v}_{\Lambda}^2  = - \frac{1}{3}  - \frac{2}{3 c} \sqrt{\Omega_{\Lambda} } \; cos y + \frac{1}{6 c} \; \sqrt{ \Omega_{\Lambda} } \left[ \frac{ \frac{1 - \Omega_{\Lambda} }{1 - \gamma a}+ \frac{2}{c} \sqrt{ \Omega_{\Lambda} } \left( 1 - \Omega_{\Lambda} \; cos^2 y \right) }{ 1 - \frac{\Omega_{\Lambda} }{c} \; cos y  } \right].
\end{equation}

\input{epsf}
\begin{figure}
\epsffile{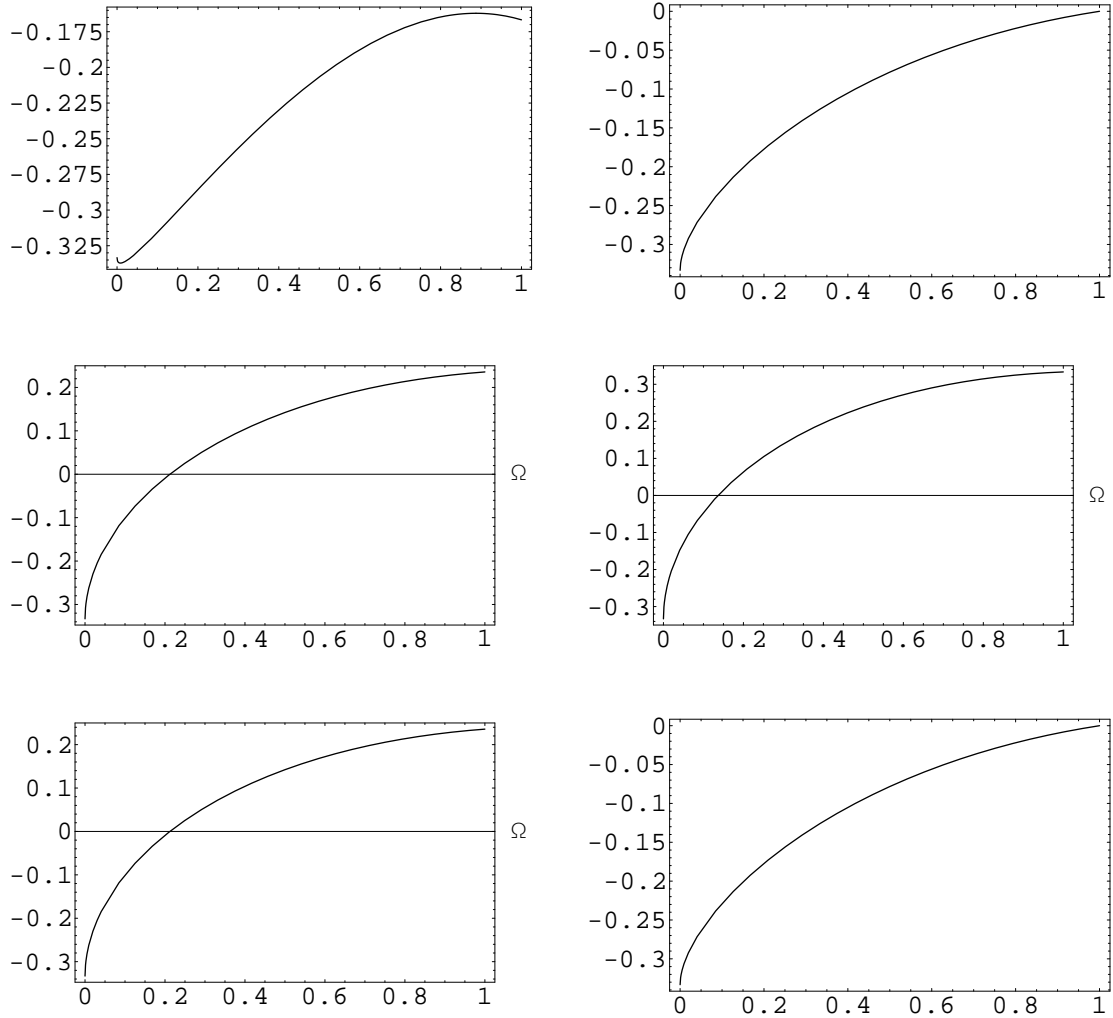}
\caption{shows the plot of ${\it v}_{\Lambda}^2$ versus $\Omega_{\Lambda} $ for different values of
$y$ with $c=1$, $\gamma = 1/3$ and $a=1$, in the first array the figures are  for $y =  \frac{\pi}{3} \; $ and $\; y =  \frac{\pi}{2} \; $, in the second array for  $ \; y =   \frac{1.5 \; \pi}{2} \; $,  $\; y =  \pi \; $ and in the third array for $y =   \frac{2.5 \; \pi}{2} \; $, $y =  \frac{3 \pi}{2}$.}
\end{figure}

The variation of ${\it v}_{\Lambda}^2$ with  $\Omega_{\Lambda}$ is shown in
 {\it fig.} 2 for different $y$ values. It is found that for a  given value
 of $c, \; a, \; \gamma$, the model admits a positive squared speed
 for $\Omega_{\Lambda} > 0$. Thus for a stable model we require
$\Omega_{\Lambda}$ positive and bounded from below.  We also note
that for
 $  \frac{(2n+1) \pi}{2} < y < \frac{(2n+3) \pi}{2}$,
 (where $n$ is an integer)   no instability develops.
 We took a few cases e.g.,  $n = 0$ in  {\it fig.} 2,  which shows that  for
 $y \leq \frac{\pi}{2}$ and $y \geq \frac{3 \pi}{2}$,  the squared speed
 for holographic dark energy  becomes negative which leds to instability.
 But,  the squared  speed is positive for the region $ \frac{\pi}{2} < y
  <  \frac{3 \pi}{2}$ with $n=0$, which implies stability.
   It is also found that for $y =0$ i.e., in flat case
 the holographic dark energy model is always unstable [30].

\section{  Discussions : }

In this paper we explored two  aspects (i) an equivalent representation of MCG with a scalar 
field and (ii) a holographic dark energy model with MCG 
 in FRW universe.
 In sec. 2 the equivalent scalar field potential
corresponding to the fluid described by  the MCG is obtained both in
a flat and in non-flat universe. We note that the potential  in {\it
Ref.} [19] for a generalised Chaplygin gas is recovered here for
$\alpha = 1$ and $B =0$. However, in a modified Chaplygin gas we
obtain a new potential determined by the parameters $A$ and $\alpha$
introduced in the equation of state. The potential asymptotically
approaches to a constant   when (i) $\phi \rightarrow 0$ and (ii)
$\phi \rightarrow \pm \infty$ for $\alpha =1$. In the non-flat case
although it is not so simple to obtain an analytic function for the
potential in terms of the field we discuss here two special cases in
which potentials are expressed as a function of the field $\phi$.  For
example, $B= \frac{1}{3}, \; \; A \neq 0$, one gets a new scalar field
potential, the shape of  the potential is similar to that one
obtains  for  a rolling tachyon [29]. It admits a small positive
effective cosmological constant at a late epoch. Thus the MCG is useful to describe
an accelerating universe at late epoch. In sec. 4 we   obtain the evolution 
of the field corresponding to  the holographic dark energy which is taken
in the form of MCG
and the corresponding potential in a non flat universe is determined.  Thus it is
important to study  a closed or open universe to account for the the
observational facts. The equation of state for the holographic dark
energy  considered by   Setare [31] in the case of generalized
Chaplygin gas is recovered here for $B = 0$ and $\alpha = 1$. We
note that in the closed model of the universe, the holographic dark
energy is stable for a given domain of the values of
$\Omega_{\Lambda}$. It is also observed that the inclusion of a
barotropic fluid in addition to the generalized Chaplygin gas (which
is MCG) does not change the shape of the potential. The evolution
of the holograpic dark energy field is determined in terms of
the  parameters $A$ and $C$, which in turn depends on the parameter
  $\alpha$ that appears in the equation of
state for MCG.  Thus it  is found that  the form of the
potential does not  change even if one considers a barotropic fluid in
addition to GCG. However, there is a change in the overall holographic dark
energy density because of the extra barotropic term in MCG.  
The holographic dark energy is found to be stable for a
restricted
 domain of the event horizon $R_{H}$ determined by $n =
 3 (1 + B)( 1 + \alpha)$
 for a positive $\Omega_{\Lambda}$ in a closed  universe, which is shown here.

\vspace{0.5in}

{\large \it Acknowledgement :}

Authors would like to  thank IUCAA  Reference Centre at North Bengal
University (NBU) and Physics Department, NBU  for providing facility
to initiate the work. BCP would like to thank Third World Academy of
Sciences $\bf (TWAS)$ for awarding Associateship to visit {\it
Institute of Theoretical Physics, Chinese Academy of Sciences,
Beijing } and UGC for awarding Minor Research Project ({\it No. F.}
32-63/2006 (SR). BCP would like to thank {\it Miao Li } for a
fruitful discussion.

\pagebreak


\begin{thebibliography}{99}

\bibitem{kn:1}  N. A. Bahcall {\it et al.}, {\it Science} {\bf 284}, 1481 (1999); W. J.
Percival {\it et al.}, {\it Mon. Not. Astron. Soc.}, {\bf 327}, 1297 (2001); M. Tegmark {\it et al.}, {\it Phys. Rev. } D {\bf 69}, 103501 (2004); L. Verde {\it et al.}, {\it Mon. Not. Astron. Soc.}, {\bf 335}, 432 (2002).

\bibitem {kn:2} D. N. Spergel  {\it et al.}, {\it Astrophys. J. Suppl. Ser.}, {\bf 148}, 175 (2003).

\bibitem {kn:3} D. J. Eisenstein   {\it et al.}, {\it Astrophys. J. } {\bf 633}, 560 (2005).

\bibitem{kn:4}  D. N. Spergel {\it et al} {\it Preprint } astro-ph/0603449 (2006);
A. G. Riess {\it  et  al},  {\it Astrophys. J.}  {\bf 607} 665 (2004);
 S. Perlmutter  { \it et  al}, {\it Astrophys.  J.}  {\bf 598} 102 (2003);
 P. de Bernardis  {\it et al},  {\it Nature} {\bf 404} 955 (2000);
 S. Perlmutter  { \it et  al}, {\it  Astrophys. J.}  {\bf 517}
 565 (1999);
  S. Perlmutter  { \it et al}, {\it  Nature }  {\bf 51} 391 (1998).

\bibitem{kn:5} A. D. Linde,  { \it Int. J. Mod. Phys.} A {\bf 17} 89 (2002);
L. Mersini,  {\it Mod. Phys. Lett.} A {\bf 16} 1933 (2001); D. H. Lyth,
D. Roberts and M. Smith, {\it Phys. Rev.} D {\bf  57} 7120 (1998);
R. H. Brandenberger, {\it Rev. Mod. Phys.}  {\bf 57} 1 (1985); A. D. Linde, {\it Rep. Prog. Phys.} {\bf 47} 925 (1984);
A. D. Linde, {\it Phys.
Lett.} B {\bf  108}  389 (1982);
 A. H. Guth, 
{\it Phys. Rev.} D {\bf  23} 347 (1981).

\bibitem{kn:6} A. A. Starobinsky, {\it Phys. Lett.} B {\bf  99} 24 (1980).

\bibitem{kn:7}  S. Mukherjee, B. C. Paul,
N. K. Dadhich,  A. Beesham and  S. D. Maharaj,  {\it Class. Quantum
Grav.}  {\bf 23} 6927 (2006); P. S.  Debnath  and B. C. Paul,
      {\it Int. J. Mod. Phys.} D {\bf  15} 189 (2006); S.
Mukherjee,  B. C. Paul,   A. Beesham and  S. D. Maharaj, {\it
Preprint}
     gr-qc/0505103 (2005); B. C. Paul  and  A. Saha,
     {\it Int. J. Mod. Phys.} D {\bf  11} 493 (2002); G. Magnano 
     and S. M. Sokolowski,  {\it Phys. Rev.} D  {\bf  50} 5039 (1994);
 B. C. Paul, D. P. Datta  and S. Mukherjee,
     {\it Mod.  Phys. Lett.} A {\bf  3} 843 (1988);
M. B. Mijic, M. S.
    Morris  and  W. Suen, {\it Phys. Rev.} D {\bf  34} 2934 (1986);
    S. Gottlober and  V. Muller, {\it Class. Quantum Grav.}
     {\bf 3} 183 (1986); A. A. Starobinsky, {\it JETP Lett.} {\bf 42} 152 (1986); L. A. Kofmann, A. D.  Linde 
  and A. A. Satrobinsky, {\it Phys. Lett.} B {\bf  157} 361 (1985);
   A. Vilenkin, {\it Phys. Rev.} D {\bf  32} 2511 (1985);  S. W.  Hawking and  J. C. Luttrell, {\it Nucl. Phys.} B {\bf  247} 250 (1984);
 R. Fabri  and  M. D. Pollock,
 {\it Phys. Lett.} B {\bf  125} 445 (1983);  A. A. Starobinsky,
  {\it Sov. Astron. Lett.} {\bf 9} 302 (1983).


\bibitem{kn:8} V. Sahni  and A. A.  Starobinsky, {\it Int. J. Mod.
Phys. } D {\bf  15} 2105 (2006); U. Alam,  V. Sahni  and  A. A.  Starobinsky,
{\it JCAP } {\bf 0406}, 008 (2004);  T. Padmanabhan, {\it Phys. Rept.}
{\bf 380}, 235 (2003); J. S.  Bagla, H. K. Jassal and  T. Padmanabhan, {\it
Phys. Rev.} D {\bf  67} 063504 (2003);  U. Alam,  V. Sahni  and  A. A. Starobinsky,
 {\it JCAP } {\bf 0304} 002 (2003); T. Padmanabhan, {\it Class.
Quantum Grav.} {\bf 19} L167 (2002); T. Padmanabhan, {\it Phys. Rev.} D
{\bf  66} 021301 (2002); V. Sahni, {\it Class. Quantum Grav.} {\bf 19},
3435 (2002);  J. S. 
 Bagla, T. Padmanabhan  and  J. V Narlikar, {\it Comments Astrophys.
} {\bf 18} 275 (1996).

\bibitem{kn:9} S. Chaplygin, {\it Sci. Mem. Moscow Univ. Math.
Phys.} {\bf 21} 1 (1904).

\bibitem{kn:10}  V. Gorini, A.  Kamenshchik,
U. Moschella and V. Pasquier, {\it Preprint}  gr-qc/0403062 (2004);
A. Y. Kamenshchik,  U. Moschella and V. Pasquier, {\it  Phys. Lett.}
{\bf B 511} 265 (2001).

\bibitem{kn:11} H. Sandvik,  M. Tegmark, M. Zaldarriaga and I.  Waga , 
{\it Phys. Rev. } D {\bf  69} 123524 (2004);  R. Bean R and O. Dore,
{\it Phys. Rev. } D {\bf 68} 023515 (2003).

\bibitem{kn:12} H. B. Sandvik,  M. Tegmark,  M. Zaldariaga and  I.  Waga, {\it Phys.
Rev. } D {\bf 69}, 123549 (2004); U. Debnath, A. Banerjee,  S. Chakraborty, 
{\it Class. Quantum. Grav.} {\bf 21}, 5609 (2004); V. Gorini, A.
Kamenshchik  and U. Moschella, {\it Phys. Rev.} D {\bf  67}
063509 (2003); U. Alam, V. Sahni, T. D. Saini  and
A. A. Starobinsky, {\it Mon. Not. Roy. Astron. Soc.} {\bf 344} 1057 (2003);
 A.  Dev,  J. S. Alcaniz and D. Jain, {\it Phys. Rev.} D {\bf  67}
023515 (2003); G. Kremer, {\it Gen. Relativ. Grav.} {\bf 35} 1459 (2003).


\bibitem{kn:13} V. Sahni, T. D. Saini,  A. A. 
Starobinsky  and  U Alam, {\it JETP Lett.} {\bf 77} 201 (2003);
M. C. Bento, O. Bertolami and  A. A. Sen, {\it Phys. Rev.} D {\bf 66}
043507 (2002).

\bibitem{kn:14}  J. C. 
Fabris, S. V. B.  Goncalves, P. E. de Souza, {\it Gen. Relativ.
Grav.} {\bf 34}, 53 (2002); N. Bilic,  G. B. Tupper,  R. D. Viollier, {\it
Phys. Lett.}, {\bf B 535}, 17 (2002);  R. Jackiw, {\it Preprint}
physics/0010042 (2000); M. Bordemann and J.  Hoppe, {\it Phys. Lett.} B
{\bf 317}, 315 (1993).
\bibitem{kn:15} S. Nojiri  and  S. D. Odintsov, {\it Phys. Rev.}  D {\bf  72},
023003 (2005); {\it Preprint} hep-th/0505215;  S. Nojiri  and  S. D. Odintsov, 
{\it Phys. Rev.} D {\bf 70}, 103522 (2005); {\it Preprint}
hep-th/0408170 (2004).


\bibitem{kn:16}  H. B. Benaoum, {\it Preprint} hep-th/0205140 (2002).

\bibitem{kn:17 } M. Szydlowski and  W. Czaja, {\it Phys. Rev. } D {\bf
69} 023506 (2004).

\bibitem{kn:18} J. D.  Barrow,
 {\it Phys. Lett.} B {\bf 235}, 40 (1990);
 J. D. Barrow, {\it Nucl. Phys.} B {\bf 310} 743 (1988).

\bibitem{kn:19} V.  Gorini, A. Kamenshchik, U. Moschella and V. Pasquier,  {\it Phys. Rev.} D {\bf  69 } 123512 (2004).

\bibitem{kn:20} D.
Bigatti and  L. Susskind, {\it Preprint} hep-th/0002044 (2000); L.
Susskind, {\it Preprint} hep-th/9901079 (1999); W. Fischler and
L. Susskind, {\it Preprint}
 hep-th/9806039 (1998).
 

\bibitem{kn:21}  R. Bousso, {\it Class. Quantum
Grav}, {\bf 17} 997 (2000); R. Bousso, {\it JHEP} {\bf 9907} 004 (1999); R. Bousso,
 {\it JHEP} {\bf 9906} 028 (1999).

\bibitem{kn:22}  S. D. H. Hsu, {\it Phys. Lett.} B {\bf  594} 13 (2004).

\bibitem{kn:23}  M. Li, {\it Phys. Lett.} B {\bf  603} 1 (2004).

\bibitem{kn:24}  J.  Zhang,  X. Zhang and H.  Liu,
 {\it Phys. Lett.} B {\bf  651} 84 (2007), {\it Preprint}  0706.1185; X.
Zhang,  {\it Phys. Lett.} B {\bf  648} 1 (2007);  M. R. Setare,
{\it Preprint} 0705.3517 (2007); N. Banerjee  and  D. Pavon, {\it
Preprint}gr-qc/0702110 (2007);  B. Chen,  M. Li  and Y. Wang, {\it Nucl.
Phys.} B {\bf 774}, 256 (2007);  M. R. Setare, {\it Preprint}
gr-qc/0610008 (2006), M. R. Setare, {\it Preprint} hep-th/ 0609069 (2006);
M. R. Setare and S.  Shafei, {\it JCAP} {\bf 0609} 011 (2006); 
 J. P. Beltran,  Almeida and J. G.  Pereirs, {\it Phys. Lett.} B {\bf 636}, 75 (2006)
; J, P, Beltran,  Almeida and  J. G. Pereirs, {\it Preprint}
gr-qc/0602103 (2006);  X. Zhang, {\it Phys. Rev. } D {\bf 74} 103505 (2006),
{\it Preprint} astro-ph/0609699; S. Nojiri  and S. D.  Odintsov,
{\it Gen. Rel. Grav.} {\bf 38} 1285 (2006); {\it Preprint}
 hep-th/0506212 (2005); Y. Gong  and Y. Z. 
Zhang, {\it Class. Quantum Grav.} {\bf 22} 4895 (2005), {\it
Preprint} hep-th/0505175; X.  Zhang and  F. Q. Wu, {\it Phys. Rev.}
D {\bf  72} 043524 (2005);   D. Pavon and W.  Zimdahl, {\it Preprint}
hep-th/0511053 (2005).

\bibitem{kn:25}  Q. G. Huang   and  M. Li,  {\it JCAP} {\bf 8} 13 (2004),
{\it Preprint}
 astro-ph/0404229  (2004).
\bibitem{kn:26}  A. G.   Cohen,  D. B. Kaplan  and  A. E. Nelson, {\it Phys. Rev. Lett.} {\bf 82} 4971 (1999).

\bibitem{kn:27} G't Hooft, {\it Preprint}  gr-qc/9310026 (1993).


\bibitem{kn:28} F. Simpson, {\it Preprint} astro-ph/0609755 (2006).

\bibitem{kn : 29} F. Leblond and  A. W. Peet, {\it JHEP}, {\bf 0304},
048 (2003); N. Lambert, H. Liu  and J. Maldacena, {\it Preprint}
hep-th/0303139 (2003); C. J. Kim, H. B. Kim,  Y. B. Kim  and  O. K. Kwon, {\it
JHEP} {\bf 0303} 008  (2003); D. A. Steer  and  F. Vernizzi, {\it Phys.
Rev. } D {\bf  70} 043527 (2004).

\bibitem{kn:30}  Y. S. Myung, {\it Preprint} gr-qc/0706.375 (2007).

\bibitem{kn:31} M. R. Setare,  {\it Phys.  Lett.} B {\bf  648} 329 (2007).



\end{thebibliography}
\end{document}